# Observation of second-harmonic generation in silicon nitride waveguides through bulk nonlinearities

Matthew W. Puckett,[1,2] Rajat Sharma,[1,2,*] Hung-Hsi Lin,[2] Muhan Yang,[2] Felipe Vallini,[2] and Yeshaiahu Fainman[2]

[1]Equal contributors
[2]Department of Electrical & Computer Engineering, University of California, San Diego, 9500 Gilman Dr., La Jolla, CA 92023, USA
[*]r8sharma@eng.ucsd.edu

**Abstract:** We present experimental results on the observation of a bulk second-order nonlinear susceptibility derived from both free-space and integrated measurements in silicon nitride. Phase-matching is achieved through dispersion engineering of the waveguide cross-section, independently revealing multiple components of the nonlinear susceptibility, namely $\chi^{(2)}_{yyy}$ and $\chi^{(2)}_{xxy}$. Additionally, we show how the generated second-harmonic signal may be actively tuned through the application of bias voltages across silicon nitride. The nonlinear material properties measured here are anticipated to allow for the practical realization of new nanophotonic devices in CMOS-compatible silicon nitride waveguides, adding to their viability for telecommunication, data communication, and optical signal processing applications.

## 1. Introduction

The realization of a CMOS-compatible platform exhibiting large optical nonlinearities with low loss in the telecommunication optical frequency regime is of great interest in the design of highly efficient modulators, switches, and wavemixers. Silicon has long been the material of choice for integrated photonics due to its prevalence in the electronics industry [1], its transparency in the near-IR wavelength regime [2], and its high third-order nonlinear susceptibility [3], but it suffers from two-photon absorption, limiting its efficiency in nonlinear wavemixing applications [4]. Moreover, silicon's centrosymmetry causes it to lack a second-order nonlinear susceptibility, disallowing efficient and linear modulation based on the Pockels effect, as well as any sort of second-order wavemixing [5]. A substantial body of research has focused on circumventing or removing this shortcoming [6-9], but in more recent years, silicon nitride has emerged as a competing material platform, boasting a wider transparency window, no two-photon absorption, and an improved ease and flexibility of fabrication [10-14].

Silicon nitride is an amorphous material and, as a result of its presumed centrosymmetry, should not exhibit any second-order nonlinearity. In free space measurements, however, silicon nitride thin films have been shown to exhibit second-harmonic generation arising from bulk nonlinearities [15]. Only an extremely small coefficient has been measured in an integrated platform to date, however, and this was attributed to interface nonlinearities [16]. If silicon nitride is to become a feasible replacement for silicon, it will be necessary for it to exhibit stronger electro-optic and wavemixing effects in a waveguide configuration than what has been shown until now.

In this paper, we report on the first experimental measurement of a bulk second-order susceptibility in silicon nitride shown through both thin film free-space measurements and on-chip, phase-matched second-harmonic generation. In the latter case, phase-matching is achieved through dispersion engineering of the waveguides' supported modes, and the powers contained at the second-harmonic wavelengths are used to determine two of the waveguides' second-order nonlinear coefficients, $\chi^{(2)}_{yyy}$ and $\chi^{(2)}_{xxy}$. We additionally demonstrate how, by applying large bias fields across our waveguides, we may modify their second-order nonlinearities through the well-known electric field-induced second-harmonic effect (EFISH). This effect is determined to have potential for the future design of highly nonlinear silicon/silicon nitride hybrid waveguides. The results shown here not only improve the understanding of silicon nitride's role in CMOS-compatible integrated photonics, but also suggest its viability for the practical realization of optical devices which rely on appreciable second- and third-order susceptibilities.

## 2. Second-Harmonic Generation in Silicon Nitride Thin Films

Prior to fabricating waveguides, we conducted transmittance second-harmonic generation measurements via the Maker fringes method to verify the existence of nonzero second-order nonlinear coefficients in silicon nitride thin films [15,17]. We deposited varying thicknesses of silicon nitride onto a fused silica substrate using an Oxford Plasma-Enhanced Chemical Vapor Depositor (PECVD), and during the deposition process, the flow rates of the three precursor gases, $SiH_4$, $NH_3$, and $N_2$, were maintained at 276, 24, and 600 sccm, respectively. Four samples were fabricated with silicon nitride thicknesses ranging from 100 nm to 500 nm in order to characterize the dependence of the generated second-harmonic signal on film thickness. We focused a pulsed Ti:Sapphire laser with a pulse duration of 150 fs, a repetition rate of 80 MHz, and an average power of 100 mW at a wavelength of 800 nm onto our thin films, which were tilted at an angle of 45 degrees with respect to the normal of the incident beam. A half-wave plate was used to control the polarization of the input pump beam, and the trasmitted signal was measured using a sequence of filters chosen to provide an OD of 12 for the pump wavelength, as illustrated in Fig. 1a. The measured s- and p-polarized second-harmonic signals are plotted as a function of the pump polarization in Fig. 1b for both 200 and 500 nm-thick films. It should be noted that a bare fused silica substrate was measured as well, and was found to generate no second-harmonic signal.

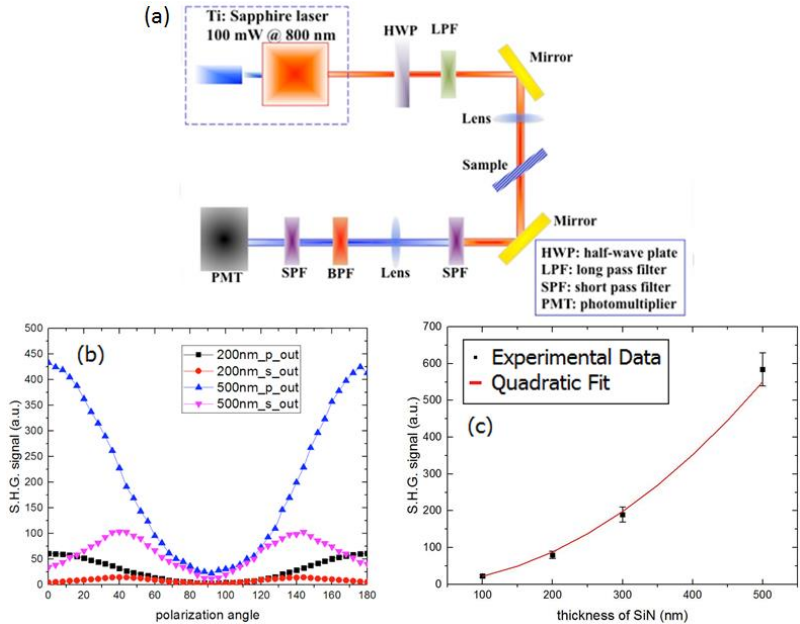

Fig. 1. (a) Schematic of the Maker fringe setup employed in this work. (b) Measured second-harmonic signal as a function of the pump polarization angle for s- and p-polarized input pump beams, and for 200 and 500 nm-thick nitride films. (c) Normalized intensity of the p-polarized second-harmonic signal (in terms of counts in the PMT) as a function of nitride film thickness, exhibiting a clear quadratic trend.

It is a well-studied property of thin film second-harmonic generation that, if a second-harmonic signal is observed, its dependence on the film thickness may be used to establish the basic nature of the film's nonlinear properties. If the film lacks a second-order susceptibility, the signal will only be generated along material interfaces where symmetry is broken, and the power contained in the second-harmonic will not vary with film thickness. If, however, the film material possesses a nonzero bulk second-order nonlinearity, the signal will vary with the square of the thickness as [15]:

$$P_{SH} \propto t_{film}^2 \qquad (1)$$

This quadratic dependence was observed, as shown in Fig. 2c, confirming the existence of a bulk nonlinearity with potential for implementation in an integrated platform. The second-harmonic signals measured in the samples were used to calculate two different components of the $\chi^{(2)}$ tensor, $\chi^{(2)}_{yyy}$ and $\chi^{(2)}_{xxy}$, assuming that silicon nitride is of the $C_{\infty,v}$ symmetry class [15] and using a modified formulation of the maker fringes method [17]. The values this treatment yielded are tabulated below wherein the coefficients calculated for different film thicknesses are found to be in very good agreement with each other, confirming the accuracy of the employed method.

**Table 1. $\chi^{(2)}$ calculated from the Maker fringes method**

|  | 100 nm | 200 nm | 500 nm |
| --- | --- | --- | --- |
| $\chi^{(2)}_{yyy}$ | 2.02 | 1.80 | 2.38 |
| $\chi^{(2)}_{xxy}$ | 0.36 | 0.30 | 0.40 |

**3. Dispersion Engineered Phase-Matching for Extraction of Second-Order Susceptibily Tensor Components**

If we assume that the guiding material of an optical waveguide has a nonzero second-order nonlinear susceptibily, the next issue to be considered in the context of wavemixing is that of phase-matching. At every position along the waveguide's propagation length, the local second-harmonic component of the nonlinear optical polarization will give rise to an electromagnetic wavelet, and the total generated second-harmonic power measured at the output of the waveguide may be considered as a superposition of these wavelets. Through coupled-mode theory, we may express the evolution of the second-harmonic mode along the waveguide as [18]:

$$\frac{dA_{2\omega}}{dz} = -j\left(\frac{\kappa}{1W}\right)\left(A_\omega(z)\right)^2 \exp(j2\Delta z) \qquad (2)$$

where $A_{2\omega}$ and $A_\omega$ are the unitless amplitudes of the second-harmonic and pump modes, respectively, $\kappa$ is the coupling coefficient (W/m), z is the position along the waveguide (m), and $\Delta$ is the phase mismatch between the two modes (rad/m). This last term is given in turn as:

$$\Delta = \beta_{0,\omega}(n_{SH} - n_P) \qquad (3)$$

where $\beta_{0,\omega}$ is the free-space wavevector at the pump frequency, and $n_{SH}$ and $n_P$ are the effective indices of the second-harmonic and pump modes, respectively. When the two effective indices in Eq. 2 are equal to one another, the phase mismatch vanishes and the waveletes emerging at the waveguide's output interfere coherently, leading to a large power in the second-harmonic mode. A separate type of phase-matching can also occur if the average of the effective indices of two separate pump modes equals that of the second-harmonic [18]. For nonzero values of $\Delta$, the second-harmonic power generated in a waveguide is a sinusoid with respect to the propagation length, whereas when Δ equals zero, it increases monotonically.

There are several methods which may be employed to achieve phase-matching. In this work we choose to modify the waveguide cross-section, leveraging its geometric anisotropy to control the effective indices of its supported modes. Let us assume a pump wavelength of 1550 nm, a corresponding second-harmonic wavelength of 775 nm, a waveguide height of 550 nm, and a sidewall angle of 83º. These values have been chosen because they correspond to fabrication results which will be shown subsequently. Using a finite element method (FEM) software such as Comsol, we may predict how the effective indices of the waveguides' supported modes change with the base waveguide width [19]. As shown in Fig. 2a, phase-matching may theoretically be attained via different components of the $\chi^{(2)}$ tensor between either (1) the TM-like pump mode and the TE- or TM-like second-harmonic mode, or (2) the combined TE- and TM-like pump modes and the TE- or TM-like second-harmonic mode. Two intersections, indicated in Fig. 2a with dashed black lines, are of particular interest because they rely on the two $\chi^{(2)}$ tensor components which were shown in the previous section to be nonzero in silicon nitride thin films. The profiles of the second-harmonic modes of interest are shown for reference in Figs. 2b and 2c.

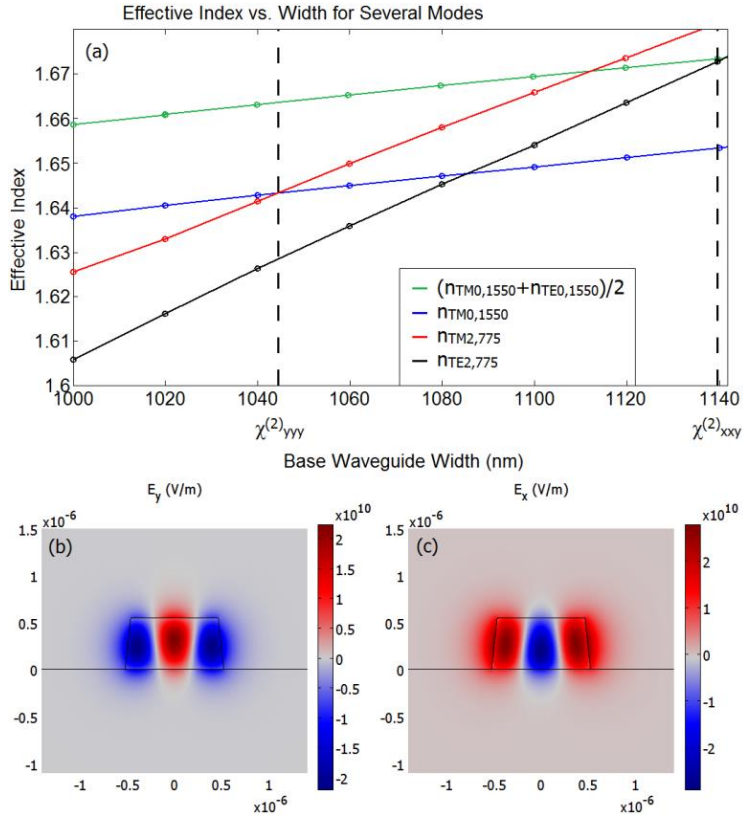

Fig. 2. (a) Effective indices of the TM-like pump (blue) and two second-harmonic modes (red, black) as a function of the waveguide width, as well as the average effective index of the TE- and TM-like pump modes (green). Phase-matching based on nonzero nonlinear coefficients occurs at widths of approximately 1045 and 1140 nm, respectively (dashed black lines). (b-c) Arbitrarily normalized electric field profiles of the TM- and TE-like second-harmonic modes, respectively, assuming a pump wavelength of 1550 nm and a base waveguide width of 1050 nm.

Coupling into the TM-like second-harmonic mode relies on the existence of $\chi^{(2)}_{yyy}$, whereas coupling into the TE-like second-harmonic is based on $\chi^{(2)}_{xxy}$. The other two intersection points shown in Fig. 2a (black/blue and red/green) may be disregarded because they rely on tensor components which are believed to be identically zero [9]. Thus, by discriminating between the orthogonal polarizations of the generated second-harmonic signal at the output of the waveguide, the two nonzero nonlinear coefficients of interest may be independently characterized.

## 4. Experimental Phase-Matched Second-Harmonic Generation

### 4.1 Fabrication

Following our free-space measurements, we set out to fabricate and characterize waveguides composed of the same nitride layer. For this particular demonstration, we began our fabrication process with silicon-on-insulator (SOI) wafers consisting of 500 nm-thick device layers and 3 μm-thick buried oxide (BOX) layers. The silicon device layer was removed through submersion in a solution of 1:10 tetramethylammonium hydroxide (TMAH) heated to a temperature of 70 ºC for one minute, and following this, we deposited 550 nm of silicon nitride, again through PECVD. In the future, nitride films may be deposited on oxidized silicon wafers, removing the need to use expensive SOI wafers [14]. After this new device layer was fabricated, we spin-coated the electron-beam resist hydrogen silsesquioxane (HSQ) onto our wafers, prebaked them for two minutes at 190 ºC, and exposed them through electron-beam lithography to patterns corresponding to 3.3 mm-long meandering waveguides with widths ranging from 700 nm to 1.12 μm in steps of 5 nm. This was done to ensure that, for at least one of the waveguides in our experiments, phase-matching would be achieved at a pump wavelength around 1550 nm. Following the lithographical step, the wafers were submerged for one minute in a solution of 1:4 TMAH to remove the unexposed HSQ. Next, the portion of the device layer left unprotected by HSQ was removed through reactive ion etching (RIE) in an Oxford Plasmalab 100. The waveguide cross-section which resulted from this fabrication process is shown in Fig. 3, and it is important to note that, although the base waveguide width matched the design width exactly, undercut during the etching process resulted in the waveguide width at the top being reduced by approximately 120 nm.

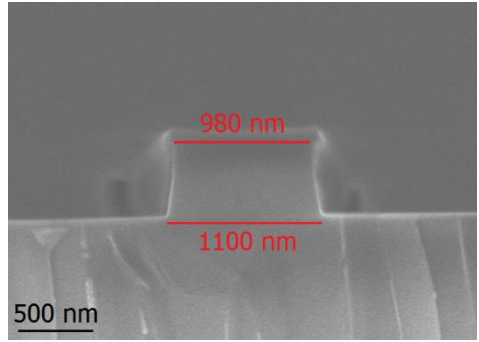

Fig. 3. SEM micrograph of an unclad silicon nitride waveguide, showing the slope of the sidewalls and the remaining unetched HSQ layer.

Following the etch step, the wafers were clad with 2 μm of silicon dioxide again using an Oxford Plasma-Enhanced Chemical Vapor Depositor. Finally, the samples were diced along the waveguide edges to allow butt-coupling via a lensed tapered fiber.

### 4.2 Characterization and Results

To measure the throughput of the pump mode, as well as any power at the second-harmonic wavelength, we employed a setup used in previous work [9,20,21]. Light from a CW laser, which was tunable from 1470 to 1570 nm, was directed into a polarization scrambler, then subsequently coupled into an erbium-doped fiber amplifier (EDFA) with a maximum output power of 1.1 W. The output of the EDFA was then directed into our silicon nitride waveguides using a lensed tapered fiber, and the transmitted light, consisting of both the fundamental and second-harmonic fields, was collected using a metallic objective. The light emitted from the objective was then magnified using two 4F lens systems and characterized by either (1) an IR powermeter, (2) a fiber-coupled spectrometer, or (3) shortpass filters and a visible spectrum powermeter. Additionally, free-space TE- and TM-polarizers were used to distinguish between different output polarizations.

Phase-matching into the TM-like mode was observed at a pump wavelength of approximately 1559 nm and a waveguide width of 1010 nm, corresponding to only a slight deviation of 35 nm from the theoretically predicted width, 1045 nm. Using the visible powermeter, we measured the wavelength- and power-dependent second-harmonic signal emitted by the waveguide, and the measured data sets are shown in Fig. 4a. In addition to the expected second-harmonic signal, appreciable optical power was measured even at wavelengths away from phase-matching, and we believe this corresponds to photoluminescence, an effect which has previously been observed in silicon nitride [22]. Nonetheless, the power in the second-harmonic could be determined by taking the difference between the non-phase-matched power and the phase-matched value. By repeating our measurement of the second-harmonic signal for several different values of pump power, we determined that the signal scaled quadratically with the pump, as expected, whereas the photoluminescence pedestal upon which the signal sat scaled linearly. This clearly indicates that the observed second-harmonic field resulted from the existence of a second-order susceptibility in silicon nitride.

The second-order nonlinear coefficient was extracted from the measured data as [18]:

$$\chi^{(2)}_{yyy} = \frac{4A_{2\omega}}{\left(\frac{\Gamma}{1W}\right)\omega\varepsilon_0 L A_\omega^2} \quad (4)$$

where $A_{2\omega}$ is the second-harmonic amplitude, $\omega$ is the angular frequency of the pump (rad/s), $\varepsilon_0$ is the permittivity of free space (F/m), L is the propagation length (m), $A_\omega$ is the pump amplitude, and $\Gamma$ is the spatial overlap between the pump and second-harmonic modes within the waveguide, defined in turn as [18]:

$$\Gamma = \iint_{SiN_x} E_{y,2\omega}^*(x,y) E_{y,\omega}^2(x,y) dxdy \quad (5)$$

where $E_{y,2\omega}$ and $E_{y,\omega}$ are the relevant electric field components of the corresponding second-harmonic and pump modes (V/m), normalized to 1 W of time-average propagating power. Again using the FEM software COMSOL, we calculated $\Gamma$ to be $7.56(10^8)$ V$^3$/m, and this gives us a $\chi^{(2)}_{yyy}$ of 0.12 pm/V. It should be noted, however, that this treatment neglects loss within the waveguide at both the pump and second-harmonic wavelengths. The loss coefficient at 1550 nm may safely be assumed to be negligibly low, using reported values for comparable fabrication processes [10], and if we assume an upper limit for the loss coefficient at 775 nm of 100 dB/cm, the extracted value of $\chi^{(2)}_{yyy}$ only increases to 0.48 pm/V [23]. We can therefore say with confidence that the actual value of $\chi^{(2)}_{yyy}$ lies between 0.12 and 0.48 pm/V.

For a waveguide width of 1070 nm and a pump wavelength of 1541 nm, we additionally observed phase-matching into the TE-like second-harmonic mode from the combined TE- and TM-like pump modes. The measured second-harmonic power is plotted as a function of the pump wavelength in Fig. 4b, and again the photoluminescence pedestal is visible away from the phase-matched wavelength. For the case of two pump modes interacting with a second-harmonic mode, Eq. 4 changes slightly to [18]:

$$\chi_{xxy}^{(2)} = \frac{2A_{2\omega}}{\left(\frac{\Gamma}{1W}\right)\omega\varepsilon_0 LA_{\omega,TE}A_{\omega,TM}} \quad (6)$$

where $A_{\omega,TE}$ and $A_{\omega,TM}$ are the amplitudes in the TE- and TM-like pump modes, respectively, and $\Gamma$ is again the confinement factor, now defined as [18]:

$$\Gamma = \iint_{SiN_x} E_{x,2\omega}^*(x,y) E_{y,\omega,TM}(x,y) E_{x,\omega,TE}(x,y) dxdy \quad (7)$$

where $E_{y,\omega,TM}$ and $E_{x,\omega,TE}$ are the relavent field components of the normalized pump modes. Using a FEM-calculated value for $\Gamma$ of 6.46($10^8$) $V^3$/m and applying a treatment identical to that used for $\chi^{(2)}_{yyy}$, we determine that the value of $\chi^{(2)}_{xxy}$ for these waveguides is between 0.06 and 0.22 pm/V. The fact that our integrated measurements yield smaller coefficients than their free-space counterparts is somewhat intuitive because, as the pump wavelength moves farther away from the band edge of any dielectric material, the value of $\chi^{(2)}$ is known to be highly dispersive [24]. The observed discrepancy between the two methods may additionally be due to the fact that, during the fabrication of the waveguides, silicon nitride is subjected to chemical, mechanical, and thermal conditions which are known to cause changes in its composition [25,26]. Optimization, including changes to the deposition conditions, may likely be made in the future to produce a higher value of $\chi^{(2)}$ in the fabricated waveguides [11,15].

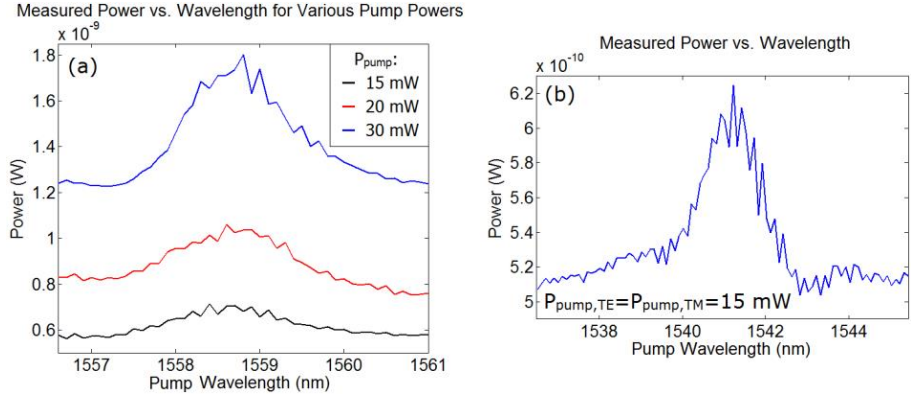

Fig. 4. (a) Measured TM second-harmonic power generated in a 1010 nm-wide silicon nitride waveguide, as a function of the pump wavelength, for several different pump powers. (b) Measured TE second-harmonic power generated in a 1070 nm-wide silicon nitride waveguide, as a function of the pump wavelength.

### 4.3 Voltage-Dependent Second-Harmonic Generation

Let us assume that a static electric field exists within a third-order nonlinear material. If an electromagnetic field propagates through the same medium, the static and oscillating fields will interact to generate a third-order polarization oscillating at twice the incident frequency, given as [27]:

$$P_{2\omega} = \frac{\varepsilon_0}{2}\chi^{(3)}(0,\omega,\omega):E_{DC}E_\omega E_\omega \quad (8)$$

where $E_\omega$ and $E_{DC}$ are the electromagnetic and bias fields, respectively. This effective second-order material polarization will then generate a new electromagnetic wave, and although this is in fact a third-order process, it mimics the behavior of second-order nonlinear materials.

This means of generating a second-harmonic electromagnetic wave is known as the EFISH effect mentioned previously.

In our silicon nitride waveguides, the standing electric field discussed above may be generated by depositing aluminum electrodes above our silicon dioxide cladding layers and applying a bias voltage vertically across the waveguides' cladding layers, as illustrated in Fig. 5a. Unlike the case of semiconductor waveguides, in which free charge carriers shield external bias fields, the field inside the nitride waveguide is relatively uniform and can be as high as $10^8$ V/m for an applied voltage of 500 V. As the voltage is increased, the field-induced $\chi^{(2)}_{eff}$ is anticipated to interact coherently with the silicon nitride's intrinsic $\chi^{(2)}$, allowing the material's total second-order susceptibility to be changed. Fig. 5b shows that the proposed fabrication has in fact been carried out, providing an SEM micrograph of the cross-section of the waveguide and electrode, and Fig. 5c shows the voltage dependence of the TM second-harmonic signal generated in the 1010 nm-wide waveguide, as measured in a fiber-coupled spectrometer. The measured increase in the second-harmonic power of 16-18% for an applied voltage of 500 V corresponds to a field-induced increase in silicon nitride's $\chi^{(2)}_{yyy}$ of approximately 8-9%. Although this enhancement to the second-order nonlinear susceptibility is small, we assert that by using materials such as silicon-rich silicon nitride, which has a third-order nonlinear susceptibility multiple orders of magnitude greater than that of typical silicon nitride [28], EFISH may eventually be used to substantially increase the conversion efficiencies attainable by material platforms similar to the one shown here.

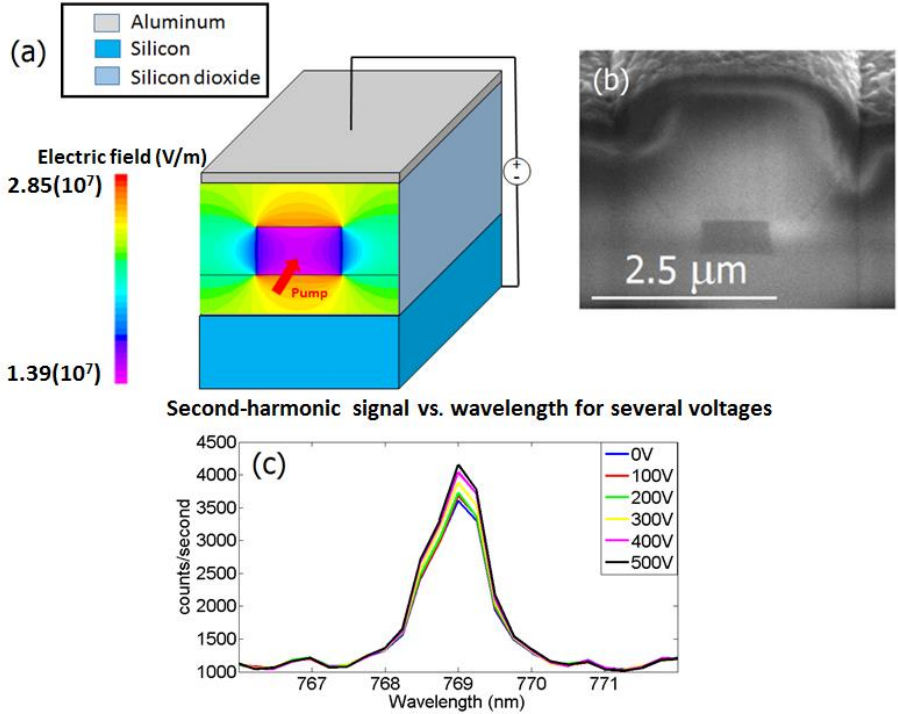

Fig. 5. (a) Schematic of the proposed mode of voltage application across a silicon nitride waveguide. Superimposed is the electric field simulated in SILVACO for the same structure with an applied voltage of 130 V. (b) SEM micrograph of a silicon nitride waveguide clad with layers of silicon dioxide and aluminum. (c) Spectrometer-measured second-harmonic signal, plotted as a function of wavelength, for several different bias voltages.

## 5. Conclusion

To summarize, we have shown for the first time that the bulk second-order nonlinear susceptibilities observed in free-space measurements of silicon nitride thin films may be used in a CMOS-compatible integrated platform to achieve second-harmonic generation which is phase-matched through dispersion engineering of the waveguide geometries. By considering the polarizations of the optical modes relevant to the wavemixing process, we characterized two specific components of the $\chi^{(2)}$ tensor, $\chi^{(2)}_{yyy}$ and $\chi^{(2)}_{xxy}$. We additionally demonstrated how the application of bias voltages across silicon nitride waveguides can lead to an effective increase in their second-order nonlinear susceptibilities, and discuss how the enhancement measured here may be improved upon in future work. As new material candidates continue to be considered for the advancement of integrated photonics, we believe characterizations like the ones presented here will be critical in determining which material platform shows the most promise for the realization of a new generation of nonlinear optical devices.

**Acknowledgements**

The authors would like to acknowledge the National Science Foundation (NSF), the Defense Advanced Research Projects Agency (DARPA), the Army Research Office (ARO), the NSF ERC CIAN, the Office of Naval Research (ONR), the Multidisciplinary University Research Initiative (MURI), and the Cymer corporation. We additionally thank UCSD's Nano3 cleanroom staff, namely Dr. Maribel Montero and Mr. Ryan Anderson, for their aassistance with sample fabrication and imaging.